%% file: main.tex
\documentclass[sigplan,screen]{acmart}

\usepackage[symbol]{footmisc}
\usepackage{setspace}
\usepackage{multirow}
\usepackage{tcolorbox} 
\usepackage{xcolor}
\usepackage{amsmath}
\usepackage{centernot}
\usepackage{alltt}
\usepackage{listings}
\usepackage{cmtt}
\usepackage[position=bottom]{subfig}
\usepackage{breakurl}
\usepackage{fancyvrb}
\usepackage{relsize}
\usepackage{soul}
\usepackage{array}
\usepackage{booktabs}
\usepackage{enumitem}
\usepackage{currfile}
\usepackage{tikz}
\usepackage{pbox}
\usepackage{xstring}
\usepackage{xspace}
\usepackage{epstopdf}
\usepackage{graphicx}
\usepackage{float}
\usepackage{makecell}
\usepackage{url}
\usepackage{algorithm}
\usepackage{algpseudocode}

\AtBeginDocument{%
  \providecommand\BibTeX{{%
    \normalfont B\kern-0.5em{\scshape i\kern-0.25em b}\kern-0.8em\TeX}}}

\algnewcommand\algorithmicforeach{\textbf{for each}}
\algdef{S}[FOR]{ForEach}[1]{\algorithmicforeach\ #1\ \algorithmicdo}
\makeatletter
\newcommand{\mybox}[1]{%
  \setbox0=\hbox{#1}%
  \setlength{\@tempdima}{\dimexpr\wd0+4pt}%
  \begin{tcolorbox}[colback={blue!50},boxrule=0.1pt,arc=2pt,
      left=1pt,right=0pt,top=1pt,bottom=1pt,boxsep=0pt,width=\@tempdima]
    #1
  \end{tcolorbox}
}
\makeatother

\makeatletter
\def\url@leostyle{%
  \@ifundefined{selectfont}{\def\UrlFont{\sf}}{\def\UrlFont{\small\ttfamily}}}
\makeatother
\floatstyle{boxed}
\newfloat{aside}{bth!}{lop}
\floatname{aside}{Aside}

 \usetikzlibrary{calc}
\makeatletter
\def\mfontsize{\f@size}

\makeatother

\makeatletter
\newcommand{\sqbox}{%
    \collectbox{%
        \@tempdima=\dimexpr\width-\totalheight\relax
        \ifdim\@tempdima<\z@
            \fbox{\hbox{\hspace{-.5\@tempdima}\BOXCONTENT\hspace{-.5\@tempdima}}}%
        \else
            \ht\collectedbox=\dimexpr\ht\collectedbox+.5\@tempdima\relax
            \dp\collectedbox=\dimexpr\dp\collectedbox+.5\@tempdima\relax
            \fbox{\BOXCONTENT}%
        \fi
    }%
}
\makeatother

\acmYear{2026}\copyrightyear{2026}
\setcopyright{cc}
\setcctype[4.0]{by}
\acmConference[SOSP '26]{Symposium on Operating Systems Principles}{September 29--October 2, 2026}{Prague, Czech Republic}
\acmBooktitle{Symposium on Operating Systems Principles (SOSP '26), September 29--October 2, 2026, Prague, Czech Republic}
\acmDOI{10.1145/3830418.3843875}
\acmISBN{979-8-4007-2585-2/26/09}

\input{macros}
\input{title}

\begin{document}
\input{abstract}
\maketitle
\pagestyle{empty}
\input{intro}
\input{bg}
\input{appmot}
\input{idea}
\input{designimpl}

\input{eval}
\input{related}
\input{conc}
\input{acks}
\input{bib}
\end{document}

%% file: macros.tex
\newcommand{\beforecaption}{\begin{spacing}{1.2}}
\newcommand{\aftercaption}{\end{spacing}}
\newcommand{\rcaption}[3]{{\beforecaption\caption{\label{#1}{\bf \small #2 } {\em #3}}\aftercaption}}

\newcommand{\crchange}[1]{{#1}}

\newcommand{\sref}[1]{\S\ref{#1}}

\newcommand{\sysname}{{Bolt}}
\newcommand{\sysnamebasic}{{BoltNaiveCF}}
\newcommand{\sysnamemetacpy}{{BoltMetaCpy}}

\newcommand{\absname}{{AgileLog}}
\newcommand{\forktext}{{cFork}}
\newcommand{\forkpluraltext}{{cForks}}

\newcommand{\fork}{{cFork}}

\newcommand{\forksever}{{sFork}}

\newcommand{\Log}{{log}}
\newcommand{\hliname}{{hierarchical log index}}
\newcommand{\hlishort}{{HLI}}

\newcommand{\ltt}{{Lazy Tail Tree}}
\newcommand{\lttshort}{{LTT}}

\newcommand{\quotes}[1]{``#1''}

%% file: title.tex
\makeatletter
\renewcommand{\@titlefont}{\LARGE\bfseries}
\makeatother
\title{\absname: A Forkable Shared Log for Agents on Data Streams}

\author{
{Shreesha G. Bhat},
{Tony Hong}, 
{Michael Noguera},
{Aishwarya Ganesan}, 
{Ramnatthan Alagappan}
}
\affiliation{%
  \institution{\emph{University of Illinois Urbana-Champaign}}
  \city{}
  \country{}
}

\renewcommand{\shortauthors}{S. G. Bhat, T. Hong, M. Noguera, A. Ganesan, R. Alagappan}
\renewcommand{\shorttitle}{\absname: A Forkable Shared Log for Agents on Data Streams}

%% file: abstract.tex
\begin{abstract}
In modern data-streaming systems, alongside traditional programs, a new type of entity has emerged that can interact with streaming data: {\em AI agents}. Unlike traditional programs, AI agents use LLM reasoning to accomplish high-level tasks specified in natural language over streaming data. Unfortunately, current streaming systems cannot fully support agents: they lack the fundamental mechanisms to avoid the performance interference caused by agentic tasks and to safely handle agentic writes. We argue that the {\em shared log}, the core abstraction underlying streaming data, must support creating {\em forks of itself}, and that such a {\em forkable shared log} serves as a great substrate for agents acting on streaming data. We propose {\absname}, a new shared log abstraction that provides novel forking primitives for agentic use cases. We design {\sysname}, a system that implements the \absname\ abstraction. \sysname\ uses many novel techniques to make forks cheap and to provide logical and performance isolation.
\end{abstract}

%% file: intro.tex
\section{Introduction} 
\label{sec-intro}

Shared logs~\cite{lazylog, corfu, boki, scalog} are the fundamental building blocks that underpin data-streaming systems~\cite{confluent2025, redpandaData2025, streamnative2019, kreps_log}. Typically, upstream sources ingest data into the shared log, which orders and makes the data durable. Downstream applications then consume the data from the shared log and process it for many purposes like real-time analytics~\cite{app-iot1}, ML inference~\cite{Waehner2024MLinf}, search~\cite{app-rtsearch1}, and others~\cite{goodhope2012building, app-fraud1}.

With the advances in reasoning capabilities of large language models (LLMs), a new kind of entity has emerged that can interact with streaming data: {\em AI agents}. Agents differ from \quotes{traditional} entities in how they operate. The behavior of a traditional application interacting with streams is defined in advance by programmers: it follows fixed logic to produce and consume records, process the consumed data, and generate outputs. In contrast, AI agents can accomplish high-level tasks on streaming data, specified in natural language. Rather than following fixed instructions, the agent uses LLM reasoning to determine on-the-fly what sequence of steps will help achieve its goal. The agent then exercises its ability to invoke external tools~\cite{schick2023toolformer, berkeley2025overlords} to execute the LLM-informed steps on streams. Today, many practical streaming systems allow AI agents to read and write streaming~data~\cite{lenses-io, confluent-stream-agents, rp-agentic-dataplane, streamnative-mcp}.

Agents interacting with streaming data unlock a variety of use cases. For example, agents allow users who do not write code to run ad-hoc data-analysis tasks on streaming data specified in natural language~\cite{rp-da-cont, rp-data-analysis, lenses-ai}. Similarly, agents can use LLM intelligence for ambiguous tasks on unstructured streaming data~\cite{streaming_agents_examples-2, streaming_agents_examples-1, streaming_agents_examples-3}. Agents can also use LLM capabilities to generate contextually relevant test events and inject them into streams for testing purposes~\cite{azraq2026eventdrivenai, lenses-ai}. 

Agents, however, introduce new problems for streaming systems. First, agents, unlike traditional tasks, are exploratory in nature: they explore many steps and paths to accomplish a task~\cite{berkeley2025overlords, giurgiu2025supporting}. Further, with the ease of building agents, many agents would operate on the system simultaneously~\cite{berkeley2025overlords, db-report}. Combined, these two aspects cause agents to place significant load, creating performance interference for other workloads, which the system must mitigate. Second, writes performed by agents can be incorrect because LLM-informed actions can be inaccurate~\cite{doltwrites, doltwrites2, berkeley2025overlords}; this causes correctness issues for applications consuming the data. Thus, the system must isolate and validate agentic writes before allowing them to take effect. Further, agents may explore many write paths and choose one. The system must thus safely allow such exploration. Finally, to enable coding and testing agents to build and test stream-processing applications, the system must provide sandboxes with realistic data. Unfortunately, today's streaming systems lack mechanisms to address these problems.

The above problems are not specific to streaming systems; agents create similar issues for any data system they interact with~\cite{berkeley2025overlords}. To address some of these problems, a few recent databases~\cite{postgres-tiger-data, neon-agent, dolt-forks}, object stores~\cite{tigris-fork}, and lakehouses~\cite{lakehouse-agent} provide {\em forking} as a core primitive that can create isolated copies of the data system on which agents can safely operate.

Inspired by these systems, we argue that {\em shared logs}, the core storage abstraction underlying streaming data, must also support creating {\em isolated forks of itself} to better support AI agents operating on streaming data. With such a forkable shared log, each agent can operate on its own fork, while traditional applications operate on the main shared log, enabling performance isolation. Forks also enable safe handling of agentic writes: an agent's writes can be isolated on a fork, and then integrated after validation into the main log. Similarly, an agent can explore many write paths in parallel on different forks, isolated from each other and the main log. Finally, each fork naturally acts as a coding and testing sandbox that contains real (production) data.

\vspace{0.05in}
\noindent
\textbf{\absname.} Based on this, we propose {\absname}, a new shared log abstraction that offers forking as a core primitive. A fork in \absname\ is a {\em cheap}, {\em logically separate}, and {\em performance-isolated} child copy of the main shared log that can be read and appended to (just like its parent), and in turn, itself be forked. Besides being the first forkable shared log, a key novelty in \absname\ is a new form of forks that we call {\em continuous forks}. 

The usual notion of forks in prior forkable data systems is that the parent and the child are disconnected from each other at the fork point: while the parent and child share the state up to the fork point, after the fork, the writes on the parent are not visible to the child and vice versa. However, this is insufficient for agents acting on real-time, streaming data. Consider an agent performing streaming analytics on a fork. For the agent to keep producing up-to-date results, the fork must continue to see the real-time data from the parent log that is ingested {\em even after the fork point}. For this purpose, \absname\ offers {\em continuous forks} (or {\forkpluraltext}). A \forktext\ shares the history with its parent up to the fork point but, importantly, {\em continuously inherits all further appends} on the parent as well.

A \forktext\ can also be appended to, but those records remain private to the fork. \forkpluraltext\ thus provide a unique form of isolation that we call {\em unidirectional write isolation}: records ingested into the parent are visible to the child but {\em not} the other way around. Thus, agents performing writes can do so on a \forktext, safely isolated from the parent, and consumers of the parent will not see those records. A key property of \forkpluraltext\ is that the records appended on a \forktext\ are linearizably interleaved with the inherited parent records. This property is critical for interleaving agentic writes on a fork with non-agentic writes on the parent in the correct order. 

In use cases like testing, agentic writes are synthetic and thus remain private to the cFork forever. However, in use cases where agents produce actual data, the writes need to be integrated into the main log. \absname\ thus provides a way to {\em promote} a \forktext\ (after validation) as the parent. Promote is also useful with exploration: agents can explore many write paths on different cForks and finally promote one. In addition to cForks, \absname\ also offers a (regular) {\em severed fork}, where the child and parent are disconnected after the fork point. Overall, these primitives provided by \absname\ help satisfy the needs of various agentic use cases over streaming data. 

\vspace{0.05in}
\noindent
\textbf{\sysname\ System.} We build \sysname, an implementation of the \absname\ abstraction. \sysname\ must satisfy two requirements. First, creating forks must be cheap and quick. Second, forks must be isolated in performance: agentic tasks on child forks should not interfere with workloads on the parent. Since forks must be cheap, any design that copies data to create a fork is a non-starter. To avoid copies, one could host forks on the same set of storage servers that host the main shared log so that the fork and the parent can share the same underlying data. This design, however, will suffer from performance interference. This points to a fundamental tension: copying data provides performance isolation but is slow, while sharing data enables fast fork creation but leads to interference.

Our architectural insight to resolve this tension is to leverage the recent \quotes{diskless} shared-log architecture~\cite{kip-1150, warpstream, confluentfreight} as the foundation for \sysname. Diskless shared logs use {\em stateless brokers} that make records durable on {\em cloud object stores} (e.g., S3) instead of local disks. A fault-tolerant {\em metadata layer} then sequences the durable records, and maintains the log metadata: an index that maps log positions to shared-storage objects. \sysname\ adopts this diskless architecture to make forks cheap and performance-isolated. Specifically, \sysname\ creates a fork with {\em zero-data-copy} by instantiating a fork's metadata to be the same as that of the parent log, making the fork point to the same objects on shared storage as the parent. \sysname\ then serves reads and appends to the fork on a separate broker; because the parent and forks are hosted on separate brokers and because cloud object stores scale well, \sysname\ cleanly isolates the performance of the parent and the forks.

While the diskless architecture provides a good base, \sysname\ must address important challenges. First, even just copying metadata to create a fork can be slow. \sysname\ makes fork creations truly low-latency by avoiding even metadata copies via a new {\em hierarchical log index} technique. Another major challenge is to implement cForks correctly and efficiently. A cFork must continuously inherit records from the parent and must linearizably interleave them with its own appends. This must be efficient: even with many cForks, there should be little impact on the parent's performance. \sysname\ implements cForks by carefully sequencing the {\em metadata of parent's new records} into the {\em child's metadata}, without data copies. \sysname\ optimizes this idea with new techniques like {\em tail-only updates} and {\em lazy-tail propagation}. The end result is that a parent log can have many cForks with little to no performance impact.

\vspace{0.05in}
\noindent
\textbf{Results.} Our experiments show that \sysname\ can create forks quickly. \sysname\ also effectively isolates workloads on the parent log from agentic tasks. \sysname's techniques help maintain the parent log's high performance even with 100s of \forktext{}s. We build three real agents powered by LLMs: an ad-hoc analytics agent for IoT data, a stream-processor testing agent, and a supply-chain-management agent. \sysname\ provides performance isolation and enables safe handling of agentic writes in these applications. In contrast, Kafka, the de facto shared log for data streaming, suffers from interference (e.g., 14$\times$ and 130$\times$ higher mean and p99 latencies), and consumer failures when agents produce problematic records.
 
\vspace{0.05in}
\noindent
\textbf{Contributions.} This paper makes four contributions. 
\begin{itemize}[noitemsep,nolistsep,topsep=0pt,parsep=0pt,partopsep=0pt,leftmargin=*]
  \setlength\itemsep{0em}
  \item We articulate how a forkable shared log serves as a better substrate for agents interacting with streaming data.  
  \item We present \absname, the first shared log abstraction to offer forking as a core primitive. \absname\ offers a novel type of fork, called a continuous fork, as well as primitives for integrating and exploring agentic writes.
  \item We build \sysname, an \absname\ implementation that realizes cheap and performance-isolated forks by adopting a diskless architecture. It uses novel techniques like {hierarchical log indexes}, {tail-only updates}, and lazy-tail propagation.
  \item We show that \absname\ benefits real agentic applications.  
\end{itemize}

%% file: bg.tex
\section{Background}
\label{sec-bg}

We describe the role of shared logs in data streaming, and explain the trend of AI agents interacting with \crchange{such systems}.

\subsection{Shared Logs and Data Streaming}
\label{bg-sl}

Shared logs~\cite{corfu, murray2025designing, boki, speclog, scalog, delos, lazylog} offer an abstraction of a durable and linearizably~\cite{linherlihy} ordered sequence of records. Shared logs expose a simple API. Clients can {\em append} records, upon which they are ordered and made durable. Clients can {\em read} records at given positions in the sequence. 

Shared logs are the core storage abstraction underlying data-streaming systems~\cite{speclog, impeller, kreps_log, kafka-sp, kleppjkreps2015kafka}. These systems support many applications like fraud monitoring~\cite{app-fraud2, app-fraud1}, real-time analytics~\cite{app-iot1} and search~\cite{app-rtsearch1}, and live inference~\cite{Waehner2024MLinf, Lin2024RealTimePredictions}. \crchange{In all these applications, upstream sources ingest data into the shared log that downstream applications consume and process}. Shared logs have gained significant attention in research and many research implementations~\cite{scalog, lazylog, corfu, boki, fuzzy} exist today. Many practical shared logs exist as well. Kafka~\cite{kafka} is perhaps the most widely used, but others~\cite{pulsar, pravega, google-sl, aws-ki-sl} are also popular.

\subsection{AI Agents Interacting with Streaming Data}
\label{bg-agents}

Recent improvement in LLMs' reasoning abilities~\cite{llm-reasoning-1, llm-reasoning-2} has enabled the widespread adoption of \emph{AI agents}. AI agents are systems that couple LLMs' reasoning with the agency to take actions, such as invoking external tools, maintaining state, and coordinating with other agents~\cite{schick2023toolformer, yao2023react}.

The ability to invoke external tools, in particular, has enabled AI agents to interact with data systems. With this ability, AI agents today can interact with databases~\cite{berkeley2025overlords, postgres-agent, postgres-tiger-data, dolt-agents}, object stores~\cite{tiger-fluid-storage, tigris-fork}, lakehouses~\cite{lakehouse-agent}, and warehouses~\cite{warehouse-agent, warehouse-agent2, dw-agent}. They can perform a variety of data tasks like retrieval, analysis, and manipulation~\cite{berkeley2025overlords, db-report, fabi-analyst-agent, calliope, powerdrill-nl-queries-pred-analysis}. Adoption of agents for data tasks is rapidly growing; for example, a recent survey shows that many companies now use agents for data tasks and that a vast majority of databases are created and operated upon by AI~agents~\cite{db-report}.

Unsurprisingly, agents have also started interacting with streaming data systems. Today, many practical systems support agentic access to streaming data~\cite{lenses-io, confluent-stream-agents, rp-agentic-dataplane}. They do so via the Model Context Protocol (MCP)~\cite{mcpmain, mcpacad}: each system provides an MCP server through which agents can read from and write to data streams~\cite{kafka-mcp-server, pulsar-mcp-1, streamnative-mcp, lenses-stream-explore, lenses-ai, redpanda_mcpmain}. 

%% file: appmot.tex
\section{Agents Acting on Data Streams}
%: Use Cases and Problems

\subsection{Real Use Cases}
\label{appmot-usecases}

AI agents operating on streaming data enable an array of use cases. \crchange{We now present these use cases, which we distilled from real scenarios that practical systems aim to support. }

\vspace{0.01in}
\noindent
\textbf{Agentic Data Analytics.} Agents can analyze streaming data, enabling non-programmers (e.g., sales personnel) to \quotes{talk} to data via natural language~\cite{fabi-analyst-agent, calliope, postgres-tiger-data}. \crchange{Agents can construct on-the-fly dashboards, answer windowed streaming queries, compute metrics over time~\cite{rp-da-cont, rp-data-analysis}, and handle ad-hoc questions~\cite{lenses-ai} like \quotes{\textit{what are the trending products in the last 24 hours?}} or \quotes{\textit{show revenue impact of the latest feature deployment}}~\cite{rp-data-analysis}. In these scenarios, an LLM helps the agent decompose tasks and iteratively determine next steps.} For example, based on LLM interaction, an agent may first probe streams to identify contents, sample records to infer schema, then read records for analysis from the target streams. Dashboards and streaming-query tasks require continuously fetching new records, while point-in-time queries require access only up to a specific point in the stream.

\vspace{0.01in}
\noindent
\textbf{Real-Time Context Access.} As many applications become agentic, they require the latest events happening in an organization to get real-time context~\cite{context-ret-2, context-ret-3, context-ret-4, context-ret-5} to make accurate decisions~\cite{rt-context-better, rt-context-better2}. For example, an agentic order-fulfillment application requires access to real-time inventory status~\cite{context-ret-2}. \crchange{As in} the previous use case, agents may iteratively probe and read streams to construct the context. 

\vspace{0.01in}
\noindent
\textbf{Agentic Stream Processing.} Stream processors are long-running programs that consume records from streams, act on them, and write outputs to other streams. Traditionally, stream processors are hand-coded (e.g., Flink jobs~\cite{flink_official} or Pulsar functions~\cite{pulsar_functions}). However, LLM-powered agents can also act as stream processors, particularly for tasks over unstructured, heterogeneous data. For example, an agent can act as a content-moderation processor to identify inappropriate images~\cite{streaming_agents_examples-2}. Here, the agent would invoke the LLM on each image from the input stream and write its reasoning and decision to an output stream. Downstream consumers (either traditional or agentic) then act on the output stream. Similar examples of agentic stream processing for customer support~\cite{streaming_agents_examples-3}, supply chain~\cite{agents_supply_chain}, and fraud-risk analysis~\cite{streaming_agents_examples-1} exist as well. Unlike prior use cases, where agents only read streams, here, they both read and write to streams.

\vspace{0.01in}
\noindent
\textbf{Agentic Coding \& Testing of Stream Applications.} Instead of having agents themselves operate on streams, one could now use AI coding agents to develop (regular) applications that interact with streams~\cite{lenses-stream-explore, streamnative-mcp, streamnative_mcp_server_video_2025}. Agents can synthesize stream processors or streaming SQL from prompts~\cite{lenses-stream-explore}. To build the application, the coding agent interacts with streams: it repeatedly reads the streams (to infer schema and understand the data) and iteratively fixes the code. Such iterative development existed even with hand-coded stream applications~\cite{netflix-data-mesh}, but the rate of such iteration will grow with agents.

Agents can also be used to test streaming applications. LLMs excel at test generation~\cite{coverup-testing, testgen-llm-meta, celik2025review} and thus agents with access to code can generate contextually relevant test cases~\cite{nvidia-test-gen}. Agentic testing is already applied to semantic testing of stream processors: the agent generates and injects test events into the stream, and checks if a stream processor behaves correctly~\cite{azraq2026eventdrivenai, lenses-ai, redpanda_mcp_tool_patterns}. For example, agents can inject synthetic transactions to test if a fraud-detection model~\cite{app-fraud1, app-fraud2} catches evolving fraud patterns~\cite{update-fraud1, update-fraud2}. Such testing can be applied to hand-coded programs or during iterative agentic development. Agents also enable schema testing, where events with new schemas are injected to check if processors can handle them. Further, agents can also perform \quotes{what-if} counterfactual testing~\cite{berkeley2025overlords}. With LLM aid~\cite{msr-what-if}, agents can do what-if tests on real-time data by simulating and injecting different event sequences and observing how downstream outputs change. Testing agents both read and write streams, but the written data is synthetic.

\subsection{Agents on Streams: Problems}
\label{appmot-problems}

\vspace{0.01in}
\noindent
\textbf{Need for Performance Isolation.} On the surface, AI agents may appear to be yet another set of clients interacting with streaming data. However, agents have characteristics that distinguish them from usual applications. As identified by recent work~\cite{berkeley2025overlords, giurgiu2025supporting}, agentic tasks are exploratory and speculative in nature: agents generate multiple courses of action to accomplish a task. This exploration may occur within a single agent or via an orchestrator spawning sub-agents, each exploring a distinct hypothesis or path. In either case, these paths can be partial attempts at the task, probes to understand the data, suboptimal solutions, or validation of prior steps. While such exploration helps improve the overall accuracy of the agent, it leads to a large number of requests being issued, potentially with suboptimal access patterns~\cite{dolt-ai-db-isolation, ana2025agent}, which places significant load on the underlying system~\cite{berkeley2025overlords, confluent_2026_predictions_2026, thenewstack_agents_databases_2025}. Given the ease of building and deploying agents, many agents will operate on a system simultaneously, multiplying this load. This will negatively interfere with traditional production applications, especially latency-critical ones. However, current streaming systems do not provide mechanisms to isolate the performance of traditional workloads from agents.

\vspace{0.01in}
\noindent
\textbf{Need for Safely Handling Agentic Writes.} Allowing agents to directly write to a data system can be risky~\cite{doltwrites, doltwrites2, dolt-agents, berkeley2025overlords}. This is because LLMs can generate inaccurate actions, causing agents to write incorrect data. This, in turn, can lead to incorrect downstream computations based on those writes. Thus, the system must isolate agentic writes from the original stream and validate them (e.g., by a human in the loop or a \quotes{reviewer} agent~\cite{dolt-ai-db, dolt-agents}). Only after such validation must the agentic writes take effect on the stream, and consumers must be able to see them. However, current streaming systems do not provide mechanisms to do this.

Further, an agent may explore many write paths either because of its inherent exploratory nature or because it uses sub-agents to explicitly explore many paths. For example, in content moderation, many sub-agents, each powered by a different LLM~\cite{streaming-agents-try}, may explore different ways to analyze the input and produce the output. It must thus be possible to explore these different paths in an isolated manner. Eventually, when a path is chosen as the desired one (e.g., based on result accuracy), it must be possible for the writes in that path to become part of the actual stream. However, today's streaming systems do not offer such support for exploration.

\vspace{0.01in}
\noindent
\textbf{Need for {\em Realistic} Sandboxes.} Coding and testing agents require sandboxes. Testing agents inject test events, which must remain isolated from the production stream. Today, agents, for this purpose, create separate streams. While this provides isolation, testing is done purely over synthetic data. However, practitioners often desire testing using real data~\cite{neon-marble, dark-canaries, feature-flags, tricentis-tip}. With streaming data, testing is most effective when executed against realistic production data which carries inherent temporal context that helps uncover corner cases. For instance, when testing a fraud detector, the agent must be able to interleave the generated fraudulent test-events within a realistic transaction stream. Similarly, coding agents can produce more accurate code with realistic data. However, today, creating such isolated yet realistic sandboxes with production streaming data is not possible.  

%% file: idea.tex
\section{A Forkable Shared Log Abstraction}
\label{sec-idea}

To better support agents operating on streaming data, we argue that the {shared log}, the core storage abstraction underlying streaming data, must support creating {\em isolated forks of itself}. \absname\ is a shared log abstraction that offers such forking capability. A fork in \absname\ is a {\em cheap}, {\em logically separate}, {\em performance-isolated} child copy of the main log.

With \absname, each agent can operate on its own fork, while traditional workloads operate on the main or root log. This provides performance isolation for workloads on the main log. \absname\ also enables safe handling of agentic writes. Writes performed by an agent on a fork are isolated from the root log, which can be validated before allowing them to take effect on the root log. Further, \absname\ enables simultaneous exploration of different write paths on separate forks. Finally, each fork acts as a sandbox with real data. 

This section first describes the \absname\ abstraction (\sref{idea-intsem}) and then how \absname\ satisfies the needs of use cases from \sref{appmot-usecases} (\sref{idea-usecases}). The next section (\sref{sec-designimpl}) describes the implementation of the \absname\ abstraction that enables \crchange{cheap fork creation} and ensures logical and performance isolation. 

\subsection{\absname\ Abstraction, Interface, Fork Semantics}
\label{idea-intsem}

An \absname\ instance has a root log, which can be forked to create child copies. Each child can be appended to and read (just like its parent), and can itself be forked. Figure~\ref{fig-interface} shows \absname's interface. \absname\ augments the traditional shared-log API (read and append) with two new calls for forking: {cFork} and {sFork}; these calls return another \absname\ instance, which can be further forked to create deeper forks. {Promote} enables writes on a fork to take effect on its parent and {squash} deletes a fork.

\noindent
\textbf{Continuous Forks.} Typically, a fork means that the parent and the child will be disconnected from each other at the fork point, and data modifications that happen either on the parent or the child will remain private to them. Forks provided by prior databases~\cite{dolt-forks, neon-branching,supabase-branching, postgres-tiger-data} offer this regular fork semantics. 

With streaming systems, live data keeps flowing in and agents need to operate on the live stream. Thus, it is insufficient to provide the regular fork, where the child stops seeing new data ingested on the parent after the fork point. Take the real-time streaming query or dashboards examples from the agentic data-analytics use case. To keep producing up-to-date results, the fork where the agent runs must continue to see the appends on the parent {\em even after the fork point}. Thus, \absname\ supports a novel form of fork that we call {\em continuous forks} or cFork. A cFork shares the history with the parent log up to the fork point, but it also {\em continuously inherits} new appends at the parent after the fork point. 

\input{fig-interface.tex} 

While appends in the parent are visible to children, children may want to insert their own records into their forks. Take the testing use case; here, the agent simulates fraud scenarios by injecting fraudulent transactions. By operating on a cFork, the agent can see the original transaction stream and also inject fraud events. However, these events must remain private to the child. \crchange{A {cFork} provides} this required isolation: writes on child forks remain private to them and consumers of the parent will not see them. Thus, cForks enable a unique form of isolation that we call {\em unidirectional write isolation}: writes on the parent are visible to the children, but not the other way around. This is in contrast to regular forks provided by prior data systems, where the isolation is~bidirectional. 

Since a cFork inherits records from its parent and also has its own records, how should these records be ordered on the cFork? \absname\ guarantees that appends on a cFork are linearizably~\cite{linherlihy} interleaved with those on the parent. That is, if an append $A$ finishes on the parent and then $B$ is appended to a cFork, then $B$ is guaranteed to appear after $A$ in the cFork.

In some use cases, appends on a cFork remain private to it forever (e.g., a \forktext\ created for testing). However, in some cases, the agentic writes must be reflected on the main log; to enable this, \absname\ allows a cFork to be promoted as the parent (as we discuss below). If such promotion is desired, then the \forktext\ must be created with the {\textit{promotable}} flag set.

Figure~\ref{fig-fork-semantics}(a) and (b) show cFork semantics. In \ref{fig-fork-semantics}(a), the cFork  acts as a live read-only copy of the parent. In \ref{fig-fork-semantics}(b), in addition, records are inserted in the live copy (e.g., $Z$ on green fork) that are linearizably interleaved with the parent records.

\noindent
\textbf{{Severed} Forks.} \absname\ also offers regular forks via a {sFork} call, where the child is \quotes{severed} from the parent after the fork point. By default, a severed fork is created from the current log tail of the parent (at the instant {sFork} is called); however, it can also be created from a past offset of the parent. In either case, a severed fork shares the parent's history up to the fork point and it stops seeing further parent updates. Some use cases may not perform writes on a severed fork as shown in \ref{fig-fork-semantics}(c), where the fork acts as a read-only snapshot of the parent. Some use cases write to a severed fork as shown in \ref{fig-fork-semantics}(d). The new records on the fork could be entirely unrelated to the parent's data (like the yellow fork in \ref{fig-fork-semantics}(d)); or, they could be some perturbations of the parent values (like the green fork).

\noindent
\textbf{Squash.} {Squash} deletes a fork and no further operations can be performed on it. For example, it can be used to dispose of unused sandboxes. \absname\ disallows \crchange{{squash}ing root logs.} 

\noindent
\textbf{Promote.} To support use cases where agentic writes must be reflected on the parent, \absname\ offers a {promote} API. A promote call on a cFork (created with the \textit{promotable} flag) of a parent makes that fork become the parent itself. Future operations that refer to the parent will \crchange{now occur} on the promoted child and ones that refer to the child are disallowed. sForks or non-promotable cForks cannot be promoted. 

{cFork} and {promote} together enable safe integration of agentic writes. \crchange{First, agentic writes can be isolated on a cFork.} The cFork can then be validated and promoted. Since \absname\ guarantees linearizable interleaving of agentic writes on the cFork with the parent log's records, \crchange{promotion preserves the correct record order. If validation fails, the cFork is squashed without affecting the main log or its consumers.}

\input{fig-fork-semantics} 

One option instead of using a cFork would be to have the agents write to a temporary log, validate the writes there, and then append to the main log. However, this allows only non-contextual, stateless checks of the current agentic writes (e.g., if the schema of the records is correct). In many cases, however, validation requires state: the current agentic writes must be validated {\em along with} the records in the main log before the fork and other records on the main log after the fork (ingested by non-agentic producers). cForks and promote enable such stateful validation. Specifically, the cFork will contain the previous records and the writes from non-agentic producers linearizably interleaved with the current agentic writes. Thus, the cFork has the required temporal context for~validation. Validation can be done, for instance, by running a copy of the stateful consumer on the cFork to check if it produces desired results. If so, the cFork is promoted; otherwise it is squashed. Even if only stateless validation is required, {cFork} and {promote} provide a neat API to validate and reflect agentic writes on the main log, instead of having to manage temporary logs. Figure~\ref{fig-fork-semantics}(e)(i) shows how {cFork} and {promote} safely handle agentic writes. The agent performs its writes on the yellow fork, which is promoted as the parent after validation.

Creating a promotable cFork introduces some restrictions on its parent. These restrictions arise because, after promotion, the parent will have additional records: ones that were inserted into the cFork. Thus, allowing reads on the parent beyond the fork point (before a promotion) could make those reads invalid (after a promotion). For this reason, \absname\ does not allow reads on the parent beyond a promotable {cFork}'s fork point until the fork is promoted or squashed. For example, in \ref{fig-fork-semantics}(e)(i), consumers on the parent cannot read entries $D$ and beyond. However, once the cFork is promoted, those entries can be read. Further, while appends to the parent can continue, \absname\ cannot return indexes for those appends beyond the promotable cFork's fork point. This is because, after promotion, the indexes would change (e.g., in \ref{fig-fork-semantics}(e)(i), $E$'s position in root will change after promotion). Not returning indexes works in practice as many applications anyway do not use them~\cite{lazylog}. Lastly, \absname\ also stops operations beyond the promotable {cFork}'s fork point on other non-promotable cForks of the parent until a decision is made. Note that these restrictions apply only when there are active promotable cForks. \crchange{Despite these restrictions, agentic applications can keep this blocking interval short by delaying creation of the promotable cFork until the proposed writes have been computed. Potentially long-running analysis, such as log reads needed to determine those writes, can be done on separate non-promotable cForks, which have no restrictions. The agent can then create a promotable cFork, append the proposed records, and perform stateful validation. Furthermore, by copying the consumer state at the fork point, validation replays only the post-fork suffix rather than the entire log, making it quick in practice.}

{cFork} and {promote} also enable agentic exploration with writes. The agent can simultaneously explore different paths on different promotable cForks and finally promote one. With many promotable cForks, only the first promote will succeed and \absname\ internally squashes others. As shown in \ref{fig-fork-semantics}(e)(ii), the agent explores two paths and promotes the yellow fork. 

\subsection{How Different Agentic Use Cases use \absname}
\label{idea-usecases}

\noindent
\underline{\textit{Agentic Data Analytics.}} Agents for streaming queries and dashboards can see the most up-to-date data by operating on a cFork of the main log as shown in Figure~\ref{fig-fork-semantics}(a). Agents for ad-hoc queries don't need the latest data and so can work on a sFork (\ref{fig-fork-semantics}(c)). By running these exploratory tasks on forks, the root log's performance would remain unaffected. 

\noindent
\underline{\textit{Real-Time Context Access.}} Agents that require fresh real-time context can obtain it via a continuous fork (\ref{fig-fork-semantics}(a)).  

\noindent
\underline{\textit{Agentic Stream Processing.}} Agentic stream processors (e.g., for content moderation) use {cFork} and {promote} to safely integrate agentic writes as shown in Figure~\ref{fig-fork-semantics}(e)(i). The agent can periodically create and write to a new cFork, validate and promote it, and repeat. If the agent explores many paths (e.g., content-moderation with different models), it can explore each path on a cFork and promote one (\ref{fig-fork-semantics}(e)(ii)). 

\crchange{\absname\ can also be used within conventional stream-processing pipelines. For example, an agent can test a candidate stream processor by creating cForks of its input streams and running the candidate against these live, isolated copies. To modify or replace a deployed processor, the application can stop the existing processor, create promotable cForks of its output streams, and direct the replacement processor's outputs to those forks. The candidate outputs and their downstream effects can then be validated before the forks are either promoted or squashed.}

\noindent
\underline{\textit{Coding and Testing Agents.}} Coding agents can operate on a (read-only) cFork (\ref{fig-fork-semantics}(a)). A testing agent can operate on a non-promotable cFork, where it can safely inject test events linearizably interleaved with the real data (\ref{fig-fork-semantics}(b)). It can explore many test cases in parallel on different cForks. For what-if testing, the agent can create many sForks to simulate alternate worlds (\ref{fig-fork-semantics}(d)). The yellow fork in \ref{fig-fork-semantics}(d) explores a completely different event sequence after the fork point. The green fork, in contrast, explores values based off of the parent's data (e.g., sensor values that are off by 5\% in an IoT stream). To do this, the agent creates a cFork, reads from it, perturbs the read values, and writes them into an sFork. 

\crchange{
\subsection{Non-Agentic Use Cases}
\label{idea-non-agentic-usecases}
While the exploratory nature of agentic applications makes the need for \absname{}'s forking primitives especially urgent, \absname\ also enables non-agentic use cases. In particular, \absname\ provides a natural substrate for testing stream processors against live production data in CI/CD pipelines. Today, developers often maintain expensive shared staging or shadow clusters that are periodically synchronized with production, or develop sandboxing and multi-tenancy mechanisms to safely share infrastructure between testing and production~\cite{kafka-test-1, kafka-test-2}. With \absname, each CI/CD test can spin off its own fork of the production stream, providing access to realistic data without requiring dedicated infrastructure. Furthermore, cForks can receive data at production rates while preserving production-workload characteristics.
}

%% file: fig-interface.tex
\lstset{
  basicstyle=\footnotesize\mttfamily,
  basewidth  = {0.48em,1.0em},
  numbers=none,
  breaklines=false,
  literate={->}{{$\rightarrow$}}1,
  keywords={interface},
  keywordstyle=\color{green!50!black}\bfseries,
  morekeywords=[3]{boolean,void,bool,Position,Record,List},
  keywordstyle=[3]\color{blue!50!black}\bfseries,
  comment=[l]{//},
  commentstyle=\color{gray},
  escapeinside={(*@}{@*)},
}

\begin{figure}[!t]
\begin{center}
\begin{lstlisting}[
  basicstyle=\fontsize{8pt}{9.5pt}\selectfont\mttfamily
]
interface (*@\textcolor{red!75!black}{\bfseries \absname}@*):
  // Traditional shared-log API calls.
  Position (*@\textcolor{blue}{\bfseries append}@*)(Record r);
  List<Record> (*@\textcolor{blue}{\bfseries read}@*)(Position from, Position to);    
  // Continuous fork; indicate if this fork is promotable.
  (*@\textcolor{red!75!black}{\bfseries \absname}@*) (*@\textcolor{blue}{\bfseries \fork}@*)(promotable = false);
  // Severed fork; optionally can fork from a past offset.
  (*@\textcolor{red!75!black}{\bfseries \absname}@*) (*@\textcolor{blue}{\bfseries \forksever}@*)(optional Position past);   
  bool (*@\textcolor{blue}{\bfseries promote}@*)(); // Promote this (promotable) cFork.  
  void (*@\textcolor{blue}{\bfseries squash}@*)(); // Delete this fork.
\end{lstlisting}
\end{center}
% \vspace{-1.3em}       
\rcaption{fig-interface}{\absname\ Interface.}{}
\end{figure}

%% file: fig-fork-semantics.tex
\begin{figure}[t]
\begin{center}
\includegraphics[scale=0.675]{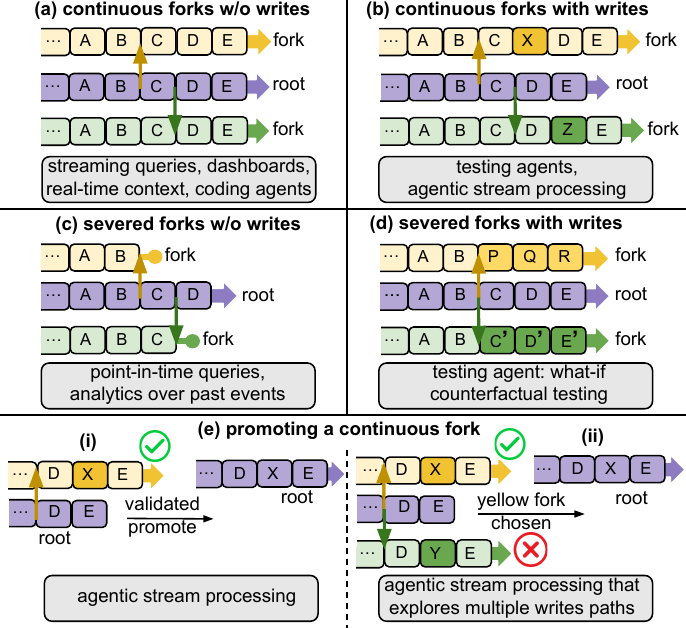}
\end{center}
% \vspace{-0.16in}
\rcaption{fig-fork-semantics}{\absname\ Primitives.}{\footnotesize The yellow (upward) and green (downward) arrows are the fork points at which the yellow and green forks were created from the (purple) root log. Writes on the forks are shown in a darker shade. Gray boxes show use cases where a particular primitive is useful.}
\end{figure}

%% file: designimpl.tex
\section{\sysname\ Design} 
\label{sec-designimpl}

\sysname\ is an implementation of the \absname\ abstraction. \sysname\ aims to satisfy a few high-level requirements. First, creating a fork must be a \emph{cheap} and \emph{low-latency} operation. Second, a fork must be \emph{isolated in performance} from the parent; this ensures that resource-intensive agentic workloads on forks do not degrade the performance of workloads on the parent log. Finally, \sysname\ must \emph{efficiently scale} to many forks.

\subsection{Design Rationale and Techniques Overview}
\label{sec-des-rationale}

One option to realize forks is to use mirroring tools~\cite{kafka-mirrormaker, uber-uReplicator, salesforce-mirus, k2k} that can continuously mirror a shared log. To create a fork, one can use these tools to mirror the parent log to the child log. Typically, these tools mirror data to a separate cluster with dedicated storage and compute. Thus, forks would be isolated in performance from the parent. However, this approach requires copying data, which makes creating and maintaining forks prohibitively slow and expensive. Another option is to build forks atop traditional shared-log designs~\cite{kafka, redpanda_kafka_broker_guide, scalog, speclog, lazylog} that use a set of replicated servers (or brokers) that store data on local disks. Here, a forked child can be hosted on the same servers as the parent and thus can share the underlying data on disk without copying. However, this would fail to provide performance isolation due to resource contention at the servers.

\vspace{0.02in}
\noindent 
\textbf{Diskless as a Substrate.} We instead realize that the recent {\em diskless} architecture in shared logs~\cite{warpstream, aiven, streamnative-ursa, kip-1150} provides a good substrate for \sysname. In contrast to traditional designs that replicate data on the local disks on a set of brokers, diskless shared logs make brokers stateless and delegate storage to scalable shared object storage (e.g., AWS S3~\cite{s3-object-store}, Azure Blobs~\cite{azureblob}). The shared-storage layer provides the required data durability. A fault-tolerant metadata layer sequences durable records\footnote[2]{While some shared logs~\cite{scalog, lazylog} also separate durability and ordering, they aren't diskless: they use local disks, not scalable cloud~stores.} and maintains an index to map log positions to shared-storage objects. The diskless architecture is increasingly popular for its two benefits~\cite{dl-benefit-1, dl-benefit-2}. First, it avoids inter-fault-domain replication traffic between brokers, which is expensive in clouds. Second, it frees operators from managing disks and helps elastically scale the system.

\sysname\ adopts the diskless architecture to make forks zero-data-copy and quick, and isolate performance (\sref{sec-overview}). At a high level, \sysname\ creates a fork by having the forked child's metadata point to the same objects on shared storage as the parent, without copying data. \crchange{Further, \sysname\ serves operations to the created fork on a separate broker from that of its parent, avoiding interference to the parent at the broker layer. Since modern cloud object stores can scale reads exceptionally well without becoming a bottleneck (e.g., with 4MB objects and 10 object-name prefixes, one can get 220 GB/s read throughput with S3~\cite{aws-s3-bp}), \sysname\ can offer performance isolation between workloads on the main log and the agents on forks.}

\vspace{0.01in}
\noindent 
\textbf{Challenges.} While diskless architectures provide a good foundation, \sysname\ must solve some critical challenges. First, how to make fork creations truly low-latency? Copying even the metadata can incur high latency. Second, how to implement cForks that continuously inherit new records ingested on the parent while linearizably interleaving appends on the child with the ones on the parent? Third, even with many cForks, how to inherit updates with little to no impact on the parent's performance? Finally, how to realize promotes? 

\vspace{0.01in}
\noindent 
\textbf{Techniques Overview.} \sysname\ addresses these challenges via various techniques. \sysname\ avoids metadata copy using a {\em hierarchical log index} that enables swift fork creations (\sref{sec-zm-fork}). \sysname\ realizes cForks as a metadata-layer operation (without data copies) by sequencing the metadata of new records on the parent into the child's metadata (\sref{sec-cf}). To scale to many cForks, \sysname\ proposes {\em tail updates} and {\em lazy-tail propagation} (\sref{sec-tech1}, \sref{sec-tech2}). Finally, \sysname\ also realizes promotes as a metadata-layer operation without data copies (\sref{sec-ps}).

\subsection{Diskless Shared Logs: A Primer}
\label{sec-primer}

Before describing \sysname, we provide a primer on diskless designs. Figure~\ref{fig-arch} shows a typical diskless shared log~\cite{warpstream, confluentfreight, streamnative-ursa, kip-1150}. The system consists of stateless brokers, a metadata layer, and cloud object storage. A single diskless instance can host {\em multiple \Log{}s} (a unit of total order), each with a unique identifier. The metadata layer is a fault-tolerant group that implements state-machine replication (SMR)~\cite{SchneiderRsm} using Paxos~\cite{lamport2001paxos} or Raft~\cite{ongaro2014search}. It maintains an {\em index} and {\em tail} for each \Log. The index maps log positions to locations of records in the object store; the tail is the next free position in the log.

\noindent\textbf{Appends.} Appends to a log can be routed to any broker (step a1 in Figure~\ref{fig-arch}). The broker batches several records from many clients (a2) and writes a large object to the object store (a3). Note that an object can contain records from different logs. Once the object write completes, the broker informs the metadata layer of the object identifier ($O$) and the metadata for each record within $O$ (i.e., the log identifier and byte ranges within $O$) (a4). The metadata layer then sequences these records by logging this metadata to its consensus log and executing them as SMR operations. Specifically, for each record in $O$ for a log $L$, \crchange{it assigns positions from $L$'s current tail}, and updates $L$'s index and tail (a5). It returns these positions (a6), and the broker acknowledges the clients~(a7).

\noindent\textbf{Reads.} Reads to a \Log\ can be routed to any broker (r1). The broker contacts the metadata layer (r2), which then looks up the index to fetch the object identifier and byte range for the requested position (r3, r4). The broker uses this metadata to retrieve the record from the store (r5, r6) and return it (r7). 

\input{fig-arch.tex}

\subsection{Forks in \sysname: Overview}
\label{sec-overview}

We now describe how \sysname\ implements the basic forking functionality atop the diskless architecture.   

\noindent
\textbf{Creating a Fork.} Conceptually, \sysname\ treats a fork as an independent \Log\ with its own total order and metadata. Upon a fork of a parent log $P$, \sysname's metadata layer creates a new child \Log\ $C$ and initializes $C$'s metadata to be the same as that of $P$. Thus, $C$'s metadata points to the same data objects on the shared storage as $P$, resulting in zero-data-copy fork creation. Concretely, when an agentic client needs a fork to operate on, it issues a fork call to any broker, and the call gets routed to the metadata layer. The fork operation is sequenced in the consensus log of the metadata layer as an SMR operation. When the operation executes, the metadata layer creates a new child \Log\ $C$, and initializes $C$'s index and tail to be the same as $P$'s index and tail at this instant. \sref{sec-zm-fork} discusses how \sysname\ does this initial fork creation efficiently to reduce latencies.  

To track the continuous-inheritance relationships between a cFork and its parent, \sysname\ maintains an {\em inheritance forest} (a collection of trees) at the metadata layer. Initially, there could be a number of {\em root} logs; these are the base logs. For each such root log, \sysname\ maintains an {\em inheritance tree}, which captures the continuous forks starting from that root. When a \fork\ $C$ is created from a root log $P$, a new tree node is created for $C$ as a child of $P$. If subsequent cForks are created from $C$, those will be added as children of $C$ in that inheritance tree. Since a severed fork does not continuously inherit updates from its parent, when a severed fork $S$ is created, a new tree is created with $S$ as the root. If cForks are created from this severed fork, those forks will be added as children of this~root. 

\sysname\ finally assigns a broker distinct from those hosting the root logs to serve the created fork to avoid performance interference. \sysname\ maintains a small pool of brokers to avoid spinning up new ones upon fork creation. The log identifier of the created fork, i.e., $C$, is returned to the client. Clients can invoke future operations on the fork using this identifier.

\noindent
\textbf{After a Fork.} After a fork is created, appends and reads to the fork $C$ are handled by the separate broker. At the metadata layer, appends to $C$ are sequenced at $C$'s current tail and added to $C$'s index and \crchange{remain logically isolated from its parent; reads are handled by reading $C$'s index.}

For severed forks, any append on the parent or the child gets sequenced only within that corresponding \Log\ index, logically isolated from each other. This is because severed forks provide bidirectional write isolation. \crchange{However, for \fork{}s, appends must not only be sequenced and recorded within the index for the \Log\ being appended to, but for all \Log{}s in its inheritance subtree.} \sref{sec-cf} describes how \sysname\ achieves this.

\subsection{Zero-Metadata-Copy Fork Creation}
\label{sec-zm-fork}

To initialize a fork's metadata, one could copy the parent's metadata. \crchange{While faster than copying data}, copying even the metadata incurs high latency and memory overhead in the metadata layer. Further, \crchange{during the metadata copy}, metadata operations on the parent's index would have to wait behind it, impacting the performance of appends on the~parent.

\crchange{\sysname\ thus avoids physically copying even metadata to create forks.} \sysname\ can do so because indexes of the fork and the parent up to the fork point would never change subsequently as logs are fundamentally {\em append-only}. Thus, \sysname\ just makes the child's index point to the parent's index for positions up to the fork point, making forks zero-metadata-copy. To realize this, \sysname\ uses a data structure that we call {\em \hliname} (or \hlishort). With \hlishort, upon a fork, \sysname\ initializes the child $C$'s index as an empty map, but $C$'s tail is initialized as the parent $P$'s tail (or as the specified offset $+1$ if a past offset is given when creating a sFork). Future appends to $C$ will be sequenced at $C$'s tail. During reads, index lookups for positions lower than the smallest position in $C$'s index are looked up in $P$. If $P$ is also a fork created from another \Log\ $Q$, it will result in a recursive lookup in $Q$'s index. \crchange{Thus, \hlishort\ avoids duplicating metadata for log prefixes shared between parents and their children.}

\subsection{Supporting Continuous Forks}
\label{sec-cf}

We now explain how \sysname\ implements cForks, where the child must see new records appended to the parent even after the fork point. \sysname\ realizes that continuous inheritance can be implemented as a {\em metadata-layer-only operation}, without any data copies. Specifically, it is sufficient to propagate the parent's metadata updates to the child's metadata. This metadata propagation indicates \crchange{the presence of new parent records to the child}. Since operations on the child also flow through the metadata layer, those operations will see the effect of the new parent records, helping establish a linearizable order across records on the child and those in the~parent.

\vspace{0.02in}
\noindent\textbf{\sysnamebasic: Naive Metadata-based \fork.} An approach to realizing \fork\ is that whenever the metadata layer sequences a record in \Log\ $P$, it synchronously updates the indexes and tails of each descendant $D\in\mathbb{D}_P$, where $\mathbb{D}_P$ is the set of all descendants in $P$'s inheritance subtree. Thus, any further append to $D$ will be linearized after $P$'s append. \crchange{We refer to this version of \sysname\ as \sysnamebasic}. Figure~\ref{fig-hli}(a) shows how \sysnamebasic\ works. Here, log $G$ is \fork{}-ed to create $R$ when $G$'s tail was 1. With zero-metadata-copy fork creation, only $R$'s tail \crchange{is set to 1}, without copying the index entries to $R$. After the fork, record $r0$ is inserted in $R$. Later, record $g1$ is inserted in $G$. This results in inserting $g1$'s index entry into both $G$'s and $R$'s indexes. Thus, when $r1$ is subsequently appended to $R$, it gets correctly linearized after $g1$. Although records in $G$ appear at different positions in $R$, by referring to the same objects on the shared storage, continuous inheritance can be realized without any data~copies.

\sysnamebasic, unfortunately, has two drawbacks. First, when an index entry $m$ is sequenced into a \Log\ $P$, it inserts $m$ into the index of every $D\in\mathbb{D}_P$. This exacerbates the memory overhead in the metadata layer ($n\times$ overhead with $n$ forks). Second, updating the index of all descendants in the critical path would impact the parent's append latency and throughput. This can be especially detrimental when agents create many cForks of a root log (e.g., during agentic~exploration). 

\subsubsection{Continuous Inheritance via Tail Updates}
\label{sec-tech1}

\input{fig-hli.tex}

To alleviate the above problems, \sysname\ makes a key observation. Upon the append of a record to a parent, the effect of the append can be reflected on the descendants by {\em updating only their \Log\ tails} and {\em not their log indexes}. Concretely, upon an append of $r$ at any $P$, \sysname\ inserts record metadata for $r$ only into $P$'s index. However, instead of also doing so for all $D \in \mathbb{D}_P$, \sysname\ only updates their tails. This tail update ensures that any future append to a descendant \crchange{$D$ will reflect the presence} of $r$, thereby achieving the effect of linearizably inserting $r$ into $D$. \sysname\ thus completely avoids any metadata duplication, i.e., each record $r$'s metadata is only inserted into one \Log\ index, \crchange{one where it was originally appended.}

\sysname\ realizes the tail-update technique by augmenting \hlishort. Figure~\ref{fig-hli}(b) shows how the augmented \hlishort\ avoids duplicating metadata even when continuously inheriting records. After the fork and insertion of $r0$ in $R$, when record $g1$ is inserted in $G$, \sysname\ avoids copying $g1$'s index entry into $R$'s index, but instead only updates $R$'s tail to 3, reflecting $g1$'s presence. Thus, when $r1$ is appended to $R$, it gets correctly linearized after $g1$ at position 3 in $R$; $R$'s tail is now 4. $g2$ is then appended to $G$ and again only $R$'s tail is updated (to 5).

\vspace{0.02in}
\noindent
\textbf{Lookups}. With tail updates, when looking up a position in a child, if the corresponding record was inherited, then it won't have an entry in the child's index (e.g., lookup for position 2 in $R$'s index in \ref{fig-hli}(b) since $g1$ was inherited). Despite this, \sysname\ must be able to retrieve the metadata for such positions. 

At a high level, the metadata for inherited records must be looked up in the parent's index. However, the challenge is that an inherited record may appear at different positions in the parent and children. In \ref{fig-hli}(b), $g1$ is at position 2 in $R$ but at 1 in $G$. \crchange{How should \sysname\ determine which position in the parent a requested position in the child maps to}? Observe that $g1$'s position in $G$ (i.e., 1) can be obtained by subtracting from $g1$'s position in $R$ (i.e., 2), the number of \emph{locally appended records} to $R$ prior to $g1$ (1 record). Thus, to look up inherited records, \hlishort\ augments the entry for a position $x$ in a \Log\ $L$'s index with another field: a cumulative count of records locally appended to $L$ up to and including $x$. 

The lookup works as follows. If a requested position $i$ is in a log $L$'s index, then \sysname\ simply returns that metadata entry. If not, \sysname\ retrieves the locally appended entries in $L$ before $i$: it finds the largest position smaller than $i$ in $L$'s index and gets the corresponding local count $l$. It then looks up $i-l$ in $L$'s parent, recursing further if $i-l$ is also an inherited position in $L$'s parent. For example, in \ref{fig-hli}(b), position 2 is not in $R$'s index. Thus, \sysname\ finds the largest position $<2$ in $R$, i.e., the 1\textsuperscript{st} record. By subtracting local count of 1 from the requested position 2, \sysname\ looks up $R$'s parent $G$'s index at position 1 which correctly returns $g1$'s metadata. In comparison, \sysnamebasic\ (\ref{fig-hli}(a)) performs a single lookup on $R$'s full index, but does so at the cost of metadata duplication.

\subsubsection{Lazy Tail Propagation}
\label{sec-tech2}

Even with tail updates, \emph{eagerly} iterating over each descendant and updating its tail in the critical path of an append to the parent would affect the performance of the parent. Instead, \sysname\ uses a {\em lazy} mechanism. Here, \sysname\ updates a descendant $D$'s tail only when there is an operation on $D$ that requires querying its current tail (i.e., an append to $D$ or a read to positions beyond the currently reflected tail in $D$).

\vspace{0.02in}
\noindent
\textbf{Efficient Tail Updates.} \sysname\ realizes tail updates and lazy propagation efficiently even with wide or deep forks using a data structure that we call \ltt\ (\lttshort). \crchange{Logically, \lttshort\ is an inheritance tree of tails, i.e., each \Log\ with its current tail forms a node, with edges between parents and inherited children. In this logical representation, an insert into a parent $p$ must not only update the parent tail node, but must also recursively update the tails of each descendant node that lies in $p$'s inheritance subtree. This requires a data structure that can efficiently encode and update subtrees. To do this, physically, we maintain} \lttshort\ as an Euler Tour of this logical tree stored in a balanced binary search tree (BST)~\cite{henzingerkingett}. An Euler Tour $E$ of a tree $T$ enumerates the nodes in the order they would be seen during a depth first traversal of $T$~\cite{euler_tour_technique}. In the Euler tour $E$, every subtree rooted at a node of $T$ appears as a contiguous range within $E$. Thus, a subtree update in $T$ converts to a range update on the BST. \sysname\ then implements range updates lazily on the BST via standard lazy-propagation techniques in logarithmic time~\cite{segment-tree-lp}. Similarly, point queries for tails on the BST take logarithmic time. Thus, forks could be arbitrarily deep or wide, but the complexity remains logarithmic in the number of \Log{}s.

\subsection{Promotes and Squash}
\label{sec-ps}

\noindent\textbf{Promote.} In \sysname, promoting a cFork can change the parent log beyond the fork point. Thus, parent reads cannot be allowed beyond it. Similarly, no appends or reads beyond this point can be allowed on any non-promotable descendant. To achieve this, \sysname\ maintains the earliest such position \textit{earliest-fp} (fork point of the earliest promotable cFork that still exists) for each log. Both direct log reads and recursive reads from non-promotable descendants that go beyond \textit{earliest-fp} are blocked. To block appends on non-promotable descendants, \sysname\ augments the lazy tail tree with the ability to block or unblock a subtree through the same lazy-propagation techniques as earlier. Further, since promotes can change the positions on the parent log, \sysname\ stops returning the append indexes after \textit{earliest-fp}.

\sysname\ implements {promote}s via metadata copies, without any data copy. To promote a child forked at position \textit{fp}, metadata entries from the child after \emph{fp} are copied over to replace entries beyond \textit{fp} in the parent. Unlike fork creation, where the history will be long and thus metadata copy can be slow, copying metadata on a promote is a reasonable design choice since only metadata entries \crchange{after \emph{fp} are copied.} 

\noindent\textbf{Squash.} Squashing a log clears the metadata-layer state for it and all its children recursively. Apart from deallocating the indexes, \sysname\ also removes the corresponding nodes from the lazy tail tree. Lastly, squashing a promotable cFork also updates the parent's \textit{earliest-fp}, and unblocks it if the squashed log is the only existing promotable cFork of its parent.        
 
\crchange{
\noindent\textbf{Lease-based Internal Squash.}
A promotable cFork imposes several restrictions on its parent, such as preventing reads beyond {\em earliest-fp}. Therefore, a rogue agent that creates a promotable cFork but neither squashes nor promotes it for an extended period can significantly degrade parent performance. To address this, \sysname\ optionally allows an authenticated entity to bound the lifetime of promotable cForks by configuring a lease duration. When leasing is enabled, \sysname's metadata layer internally squashes any promotable cFork whose lifetime exceeds the configured duration. This mechanism limits the impact of rogue or misbehaving agents.
}
\crchange{
\subsection{Alternative: Tagged Global Address Space Design}
\label{sec-alt-design} 
In \sysname, we model each root and fork as an independent log with its own local address space. An alternative design could place all roots and forks in a single global address space, using record \emph{tags} to identify the logs to which each record belongs. Prior systems have similarly partitioned a global address space among multiple logs~\cite{vcorfu, boki}. Because global positions remain stable when records are reassigned during promotion, this design could return append positions even while promotable cForks exist, removing one restriction of our promote implementation (see~\sref{idea-intsem}). It would not, however, eliminate the restriction on parent reads beyond the fork point. Moreover, efficient reads would still require per-log indexes, while a global address space would expose non-contiguous offsets to applications and impose an order over otherwise unrelated records. These tradeoffs illustrate that the \absname\ abstraction is independent of any particular address-space organization: \sysname\ implements one point in the design space, while alternative implementations can optimize for different application requirements. }

\subsection{Implementation}
\label{sec-impl}

We implement \sysname\ in C++ (\textasciitilde{}21K LOC). We use eRPC~\cite{erpc} for communication and MinIO~\cite{minio} as the shared-storage layer, but any S3-compatible store works. The metadata layer uses a Raft implementation~\cite{willemt-raft}. We equip brokers with a local object cache to improve \crchange{read performance}. While \sysname\ hosts forks on separate brokers from those of their parents for performance isolation, it co-locates many forks of a parent on the same broker for two reasons. \crchange{First, these forks often read the same inherited objects (e.g., across many analytics agents); co-location allows them to reuse broker-local cached copies and reduce read traffic to the shared-storage layer}. Second, grouping many forks to a broker reduces the overhead on the metadata layer. That said, applications \crchange{can configure \sysname\ to host forks on separate brokers.}

\input{eval/append-fork-cts/fig-append-fork-cts}

%% file: fig-arch.tex
\begin{figure}[t]
\begin{center}
\includegraphics[scale=0.8]{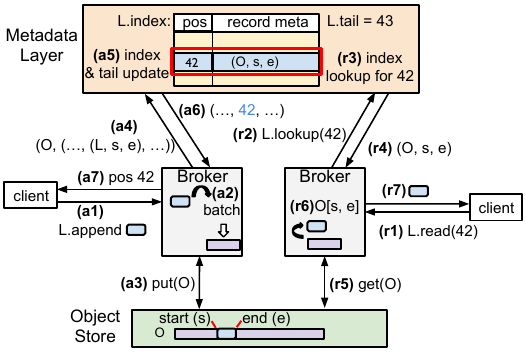}
\end{center}
% \vspace{-0.18in}
\rcaption{fig-arch}{Diskless Primer.}{\footnotesize The figure shows how appends and reads work (for entries in log L). Only $L$'s metadata is shown in the metadata layer.}
\end{figure}

%% file: fig-hli.tex
\begin{figure}[t]
\begin{center}
\includegraphics[scale=0.9]{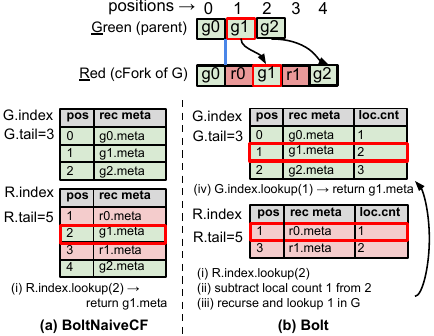}
\end{center}
% \vspace{-0.2in}
\rcaption{fig-hli}{Continuous Inheritance.}{\footnotesize $G$ is \fork{}ed to get $R$ (blue line is the fork point). (a) and (b) show how \sysnamebasic\ and \sysname\ implement cForks, respectively. Lookup of position 2 in $R$ is also shown (red boxes).}
\end{figure}

%% file: eval/append-fork-cts/fig-append-fork-cts.tex
\begin{figure}
\centering
\begin{minipage}{0.47\columnwidth}
\centering
\includegraphics[width=\textwidth]{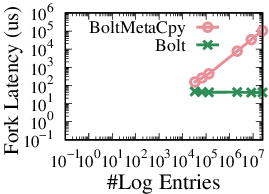}
\rcaption{fig-fork-lat}{Fork Latency. }{\footnotesize}
\end{minipage}
\hspace{0.01in}
\rule[-1.5cm]{0.2mm}{3.6cm}
\hspace{0.01in}
\begin{minipage}{0.47\columnwidth}
\centering
\includegraphics[width=\textwidth]{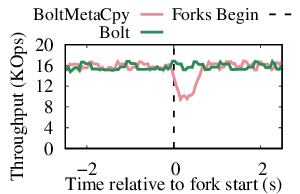}
\rcaption{fig-append-fork-cts}{Parent Throughput during Fork Creation. }{\footnotesize}
\end{minipage}
\vspace{-0.2in}
\end{figure}

\begin{figure*}
\begin{minipage}{0.4\textwidth}
\centering
\subfloat[\crchange{Isolation from Reads}]{\includegraphics[width=0.5\textwidth]{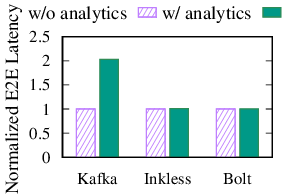}}
\subfloat[\sysname: Isolation from Appends]{\includegraphics[width=0.5\textwidth]{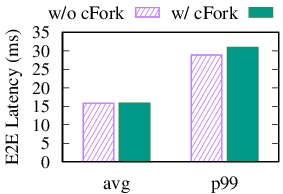}}
\rcaption{fig-noisy-neighbour}{Performance Isolation with \sysname. }{\footnotesize}
\end{minipage}
\hspace{0.01in}
\rule[-1.6cm]{0.2mm}{3.7cm}
\hspace{0.01in}
\begin{minipage}{0.36\textwidth}
\centering
\includegraphics[width=\textwidth]{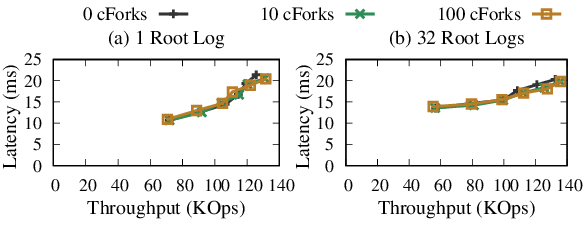}
\rcaption{fig-append-fork}{Performance Impact on Parent with many cForks.}{\footnotesize}
\end{minipage}
\hspace{0.01in}
\rule[-1.6cm]{0.2mm}{3.7cm}
\hspace{0.01in}
\begin{minipage}{0.18\textwidth}
\centering
\includegraphics[width=\textwidth]{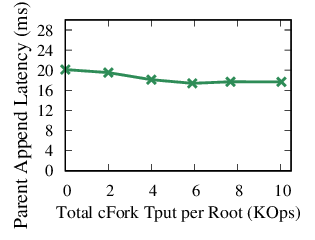}
\rcaption{fig-append-fork-loaded}{\crchange{Parent Latency Impact with cFork Append Load.}}{\footnotesize}
\end{minipage}
\vspace{-0.2in}
\end{figure*}

%% file: eval.tex
\section{Evaluation}
\label{sec-eval}

Our evaluation answers the following questions:
\begin{itemize}[noitemsep,nolistsep,topsep=0pt,parsep=0pt,partopsep=0pt,leftmargin=*]
\setlength\itemsep{0em}
\item Does \sysname\ enable low-latency fork creation? (\sref{sec-eval-fork-lat}) 
\item Does \sysname\ isolate the performance of workloads on root logs from those running on their forks? (\sref{sec-eval-perf-iso})
\item Can \sysname\ support many cForks without impacting the performance of workloads running on the parents? (\sref{sec-eval-fork-scale})

\item How do the techniques to implement cForks help? (\sref{sec-eval-techniques})
\item What is the metadata memory overhead in \sysname? (\sref{sec-eval-mem-over})
\item What is the overhead of \hlishort's recursive lookups? (\sref{sec-lookup-overhead}) 
\item \crchange{How does complex nesting impact parent inserts? (\sref{sec-nesting-impact})}
\item Are promotes of cForks fast in \sysname? (\sref{sec-eval-promote-alt})
\item Does \sysname\ benefit real agentic applications? (\sref{sec-eval-real-apps})
\item \crchange{How does \sysname\ deal with rogue agents? ({\sref{sec-eval-rogue-agents}}})
\end{itemize} 

\noindent\textbf{Setup.} We experiment on an xl170~\cite{cloudhardware} CloudLab~\cite{cloudlab} cluster. We use nine nodes for MinIO and three replicas in the metadata layer. Each broker runs on its own node. We use 4KB records. Since no existing shared log offers forks, we mostly compare \sysname\ to its many variants. In some experiments, we compare to Kafka, a popular shared log for data-streaming, and Inkless~\cite{aiven}, Aiven's fork of Kafka that implements the diskless shared-log architecture.

\subsection{Fork Creation Latencies}
\label{sec-eval-fork-lat}
\input{eval/append-fork/fig-append-fork}

We first measure fork latency. We compare \sysname\ against \sysnamemetacpy, a variant that creates forks by copying the metadata of the parent to instantiate the child index. Figure~\ref{fig-fork-lat} shows fork latency with varying lengths of the parent log. In both systems, the created forks are assigned to a pre-existing broker different than that of the parent. By avoiding any data or metadata copy, \sysname\ has low latencies (\textasciitilde{}50us), regardless of the log length. Although \sysnamemetacpy\ avoids data copies, it incurs high latency to copy metadata (e.g., 100ms with 25M entries, three orders of magnitude slower than \sysname). 

Copying metadata not only hurts fork latency but also the parent's performance. This is because record appends to the parent must sequence the metadata into the parent's index, which is affected by the copy to create the fork. Figure~\ref{fig-append-fork-cts} shows this: during the creation of 100 forks (\crchange{such as} at the start of an agentic exploration), with \sysnamemetacpy, the parent's throughput suffers \crchange{during fork creation}; in contrast, with \sysname, the parent's performance~remains~unaffected.  

\subsection{Performance Isolation}
\label{sec-eval-perf-iso}

We next show that \sysname{} provides performance isolation for a latency-critical (lc) workload from a read-heavy workload resembling agentic data analysis. The lc-workload appends and reads records at the tail. The analysis workload reads records in bulk from the same log (up to an offset). \crchange{We run these workloads on three systems, disk-based Kafka, Inkless~\cite{aiven} (a diskless Kafka implementation), and \sysname. Inkless does not offer any notion of isolated forks, but we can {\em simulate an isolated, read-only fork} by creating a new diskless broker that accesses the production topic-partition's data through the shared object storage layer. Disk-based Kafka neither provides any notion of forks nor a means to create diskless broker instances and hence the analysis task must share data by running on the same storage brokers as the lc-workload. In \sysname, the analysis task creates and runs on a \text{sFork} that uses its own broker but shares the parent's data via shared storage. Figure~\ref{fig-noisy-neighbour}(a) shows the end-to-end (e2e) latency of the lc-workload with and without the analysis task running alongside it. In both Inkless and \sysname, the lc-workload experiences no interference from the analysis task. This demonstrates that the diskless architecture provides a strong substrate for implementing the \absname\ abstraction. In contrast, with Disk-based Kafka, the lc-workload's latency degrades by 2$\times$ due to resource contention at the shared broker. We note that this behavior is not specific to Kafka but to any shared log with stateful brokers or storage servers that store data on local disks~\cite{speclog, lazylog, scalog}.}

We next show that the lc-workload is performance-isolated even with appends to the forks. \crchange{We do not compare to Kafka or Inkless, since they have no notion of forks that allow appends that are logically isolated from the root log.} The lc-workload appends to the root at 13Kops/s. We create a cFork that also appends its own records at 13Kops/s. Figure~\ref{fig-noisy-neighbour}(b) shows the lc-workload's e2e latency with and without the cFork. Even with appends on the fork, the mean and p99 e2e latencies are unaffected.

\subsection{Performance of Parent with Many cForks}
\label{sec-eval-fork-scale}

We now examine if the root log remains performance-isolated even when it has many cForks. We run an append-only workload on the root with 0, 10, and 100 cForks, and plot the root log's append throughput vs. latency. As shown in Figure~\ref{fig-append-fork}(a), even with many cForks which must continuously inherit appends from the root, the root log shows no performance degradation relative to the case with no forks. Figure~\ref{fig-append-fork}(b) shows the case where the system hosts 32 root logs (mimicking real deployments where many topic-partitions are hosted on a diskless instance). We then create 0, 10, and 100 cForks per root log. Even with many root logs, each \crchange{with} many forks, root logs' performance does not degrade in \sysname.

\crchange{We next evaluate whether root logs remain performance-isolated even when many of their cForks are also being appended to. We append to 8 root logs at 20Kops/s in aggregate. We then vary the number of cForks per root log with a load of 40 ops/s per cFork (resembling agentic workloads, where appends are interspersed with slow reasoning steps). Figure~\ref{fig-append-fork-loaded} shows that root log append latency remains constant as the total cFork load per root log increases from 2Kops/s with 50 cForks to 10Kops/s with 250 cForks.}

\subsection{cFork Techniques Ablation}
\label{sec-eval-techniques}

We now evaluate the different techniques that \sysname\ uses in the metadata layer to implement cForks. We compare \sysname\ with (i) \sysnamebasic, which copies metadata entries from the parent's index to the child indexes on every parent append; (ii) \sysname-ET (eager-tail), which improves over \sysnamebasic\ by updating only the tails but does so eagerly and without the lazy tail tree (LTT). The difference in these variants emerges only when the data-layer throughput is high enough to stress the metadata layer. However, producing this metadata load would require more brokers and MinIO storage nodes than we have machines. Thus, we measure the metadata-layer throughput in isolation. We emulate two sets of brokers sending requests to the metadata layer: \crchange{one set hosting the root log, which append records to the log; another set that hosts the cForks, which perform reads on the cForks.}

Figure~\ref{fig-metadata-fork} shows the root log's metadata performance with 10, 100, and 1000 cForks. First, \sysnamebasic\ performs well with 10 cForks, but its approach of copying the index entries to all children becomes a bottleneck with 100 and 1000 forks, resulting in low metadata-layer performance. Once the metadata layer bottlenecks, the overall system's performance would not scale further. Second, \sysname-ET's tail-only updates help it closely match the performance of \sysname\ for 10 and 100 forks. However, with 1000 forks, eagerly updating even just the tail becomes the bottleneck. In contrast, \sysname's metadata layer continues to scale well even with 1000 forks. This is because, unlike \sysname-ET, \sysname\ updates tails lazily (only upon a tail read on the cFork) and efficiently (via LTT). 

\subsection{Metadata Memory Overhead} 
\label{sec-eval-mem-over}

We now examine the metadata memory overhead of \sysname\ and \sysnamebasic. We measure the memory occupied to support 1000 cForks of a root log to which 1M records are appended. \sysnamebasic\ requires 4.4GB to maintain the cForks. In contrast, \sysname\ avoids metadata duplication, requiring only 8MB. \sysname's techniques thus are not only critical for high metadata-layer performance but also to reduce memory overhead.
\input{eval/metadata-lazy/fig-metadata-lazy}

\subsection{Recursive Metadata Lookup Overhead}
\label{sec-lookup-overhead}

With \hlishort\ augmented with tail updates, \sysname\ may require recursive lookups to find the metadata for a given log position (Figure~\ref{fig-hli}(b)). We measure this overhead and compare it to \sysnamebasic\ that doesn't require recursive lookups. We create nested cForks with 1M records per level and query the deepest cFork, forcing recursion to the root. Figure~\ref{fig-meta-query} plots the metadata lookup latency against nesting depth. \sysname\ has minimal degradation: only 5.2\% slower at depth 7 ($\leq 2\mu{}s$ overhead). Since this occurs in the metadata layer, which constitutes only a small fraction of the client-perceived latencies (which are in $ms$), the end impact is negligible. 

\crchange{
\subsection{Impact of Complex Nested cForks}
\label{sec-nesting-impact}

We now evaluate parent performance under complex nesting. We vary the total number of logs from 100 to 1M and, at each scale, organize the logs into four types of cFork hierarchies. The \emph{star} arrangement places every cFork directly below the root, producing a wide, single-level hierarchy. The \emph{balanced} hierarchy uses a fan-out of four, increasing both breadth and depth as it grows. The \emph{chain} represents the deepest extreme, with every cFork created from the preceding cFork. Finally, the \emph{irregular} hierarchy selects parents at random, producing uneven branching and depth. We measure the \hliname\ data structure with tail-only updates and lazy-tail propagation in isolation. Each measured root insert is interspersed with an insert and a read on distinct, randomly selected cForks. Figure~\ref{fig-branching-index} shows that hierarchy shape has little effect: for a fixed number of logs, mean insert latency on the parent differs by less than 0.5\,\textmu s across all structures. Thus, performance impact on root logs is minimal even with extreme nesting and many~cForks. Further, as the number of logs increases, latency grows only logarithmically.  
}

\subsection{Performance of Promote}
\label{sec-eval-promote-alt}
We now evaluate {promote}s. As figure~\ref{fig-promote} shows, metadata-copy-based promote incurs only 10-100s of $\mu{}s$, even with many records. In contrast, a temporary-log-based approach that copies data to promote (described in~\sref{idea-intsem}) with Kafka (\quotes{DC (Kafka)}), is much slower than \sysname. Thus, apart \crchange{from functionality} limitations (i.e., no stateful validation, no linearizable interleaving of agentic and non-agentic records) the data-copy-based approach also suffers from high latencies.

\subsection{Evaluation with Real Agentic Applications}
\label{sec-eval-real-apps}

We demonstrate \sysname{}'s benefits for real agentic tasks by building three agentic applications: an ad-hoc analytics agent for IoT sensor data, a stream processor testing agent, and a supply-chain restocking agent. Together, these applications exercise all \absname\ interface calls. All applications follow a single-agent architecture, and are built atop OpenCode~\cite{opencode} and use the Gemini-2.5-Pro LLM~\cite{gemini-2.5} for reasoning.

\noindent\textbf{Analytics Agent ({sFork}).} We build an agent to perform ad-hoc analysis of an IoT stream that has records with different metrics (temperature, humidity) and sensor status. The agent is given an open-ended task: \quotes{look for anomalies in the first 1M records}. The agent is provided a tool to read records from the shared log and a SQL tool~\cite{duckdb} to analyze the read data. The agent can issue multiple parallel tool calls, each investigating a different hypothesis. We pre-populate 1M records in the stream with some anomalous values. The agent runs alongside a latency-critical (lc) workload that monitors newly ingested sensor readings. With \sysname, \crchange{the agent creates an sFork to read records from. With Kafka, the agent directly} queries the data stream. We find that the agent \crchange{can} form complex hypotheses, discover injected spikes, and correlate them with metrics. For determinism, we capture one run of the agent and replay the trace on both systems. 

Figure~\ref{fig-analytics}(a) and (b) show the lc-workload's e2e latencies. The shaded regions are the periods where the agent reads the stream, with annotations ($nQ$) indicating $n$ parallel investigations; the non-shaded regions are the thinking phases, where the agent interacts with the LLM to evaluate results and plan next steps. With Kafka, the lc-workload suffers from latency spikes during query execution due to contention. In \sysname, lc-workload experiences no interference. Figure~\ref{fig-analytics}(c) shows the latencies during the two phases. In Kafka, in the query-execution phase, the lc-workload experiences 14$\times$ and 130$\times$ higher mean and p99 latencies, respectively, compared to the thinking phase. \sysname\ has no interference.

\input{eval/adhoc-analytics/fig-analytics}
\input{eval/promote-validate/fig-apps}

\noindent\textbf{Stream Processor Testing Agent (non-promotable {cFork}). } We build a stream processor that performs aggregates in tumbling windows~\cite{quix-window-type} of 5ms. We task an agent with testing this stream processor under corner cases such as late, malformed, and duplicate records, while running alongside a real-time workload. The agent is provided a tool that can create non-promotable {cFork}s of the root log, inject agent-generated test records into them, run the stream processor on the {cFork}, {squash} the {cFork}, and output execution traces (errors/failures). The agent can invoke this tool in parallel to run multiple tests concurrently. We note that this agent successfully triggers many buggy scenarios. Figure~\ref{fig-stream-test} plots e2e latency of the real-time workload with agentic tests running alongside. Shaded regions are test-execution phases (with annotation showing number of parallel tests); non-shaded regions are thinking phases. Despite tests running alongside, the real-time workload experiences no latency degradation. This shows that {cForks} provide \crchange{a sandboxed environment with real data without impacting workloads on the root log}.

\noindent\textbf{Supply-Chain Agent (cFork + Promote/Squash).} We build a supply-chain agent~\cite{agents_supply_chain}. The supply-chain stream contains order events for items and restock events that replenish supply. The order events are ingested by non-agentic producers. Based on historical trends, the agent proactively issues restock events for high-demand items. This stream is processed by a non-agentic downstream application. In \sysname, the agent is \crchange{given a tool} that creates a \textit{promotable} {cFork}, adds restock events into it, and validates it by running a {stateful copy} of the downstream application on it, before promoting or squashing the cFork. In Kafka, the agent directly writes restock events to the main stream. To simulate an agent mistake, we inject schema errors into an agent-created event. 

Figure~\ref{fig-promote-validate} plots the downstream processor's throughput. In Kafka, an agent mistake crashes the consumer application. \sysname\ \crchange{avoids crashes} through the promotable {cFork}, which helps safely validate agentic writes before incorporating them into the main stream. Since \sysname\ prevents reads from the main stream when a promotable {cFork} exists, there are throughput dips. However, once the {cFork} is either promoted or squashed, it unblocks the downstream consumer, allowing it to quickly catch up. Optionally, once the agent is thoroughly vetted, it can be allowed to directly write to the main stream and avoid the throughput dips. \crchange{Promotable {cFork}s enable the safe deployment of agents that write to production streams without worrying about application failures.}

\crchange{
\subsection{Rogue Agents}
\label{sec-eval-rogue-agents}
We now evaluate how \sysname\ behaves with rogue agents. In this experiment, we simulate a rogue agent that blocks forever after creating a promotable cFork. Figure~\ref{fig-rogue-agent}(a) shows that without \sysname's lease-based internal squash mechanism (see \sref{sec-ps}), such an agent can permanently affect the throughput of a consumer on the root log since reads beyond the fork point are blocked. In figure~\ref{fig-rogue-agent}(b), we see that while consumer throughput initially drops, \sysname's lease-based mechanism squashes promotable cForks internally after a configurable timeout (in this case 500ms) and parent consumer throughput heals. This prevents permanent harm to production workloads from misbehaving agents.}

%% file: eval/append-fork/fig-append-fork.tex
% Intentionally empty; figure definitions are included elsewhere.

%% file: eval/metadata-lazy/fig-metadata-lazy.tex
\begin{figure*}
\centering
\begin{minipage}{0.7\textwidth}
\includegraphics[width=\textwidth]{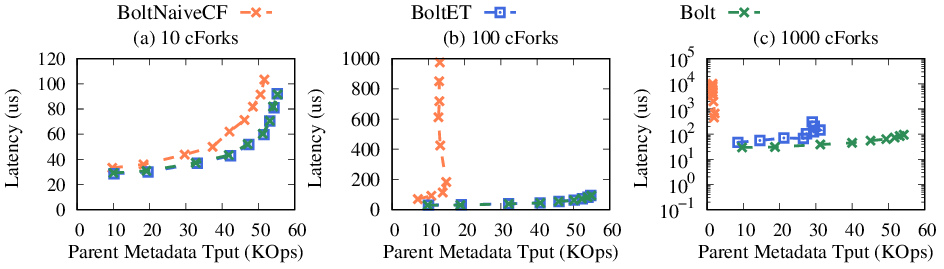}
\vspace{-0.2in}
\rcaption{fig-metadata-fork}{cFork Techniques Ablation. }{\footnotesize}
\end{minipage}
\hspace{0.01in}
\rule[-1.7cm]{0.2mm}{4cm}
\hspace{0.01in}
\begin{minipage}{0.25\textwidth}
\centering
\includegraphics[width=\textwidth]{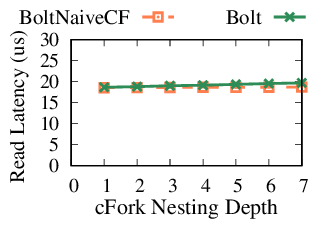}
\vspace{-0.2in}
\rcaption{fig-meta-query}{Metadata Lookup vs. cFork Depth.}{\footnotesize}
\end{minipage}
\par\vspace{-0.2in}
\end{figure*}

\begin{figure}[!t]
\begin{minipage}{0.47\columnwidth}
\centering
\includegraphics[width=\textwidth]{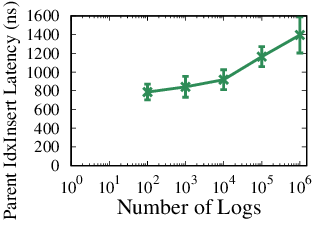}
\rcaption{fig-branching-index}{\crchange{Impact of Complex Nested cForks.}}{\footnotesize}
\end{minipage}
\hspace{0.01in}
\rule[-1.5cm]{0.2mm}{3.6cm}
\hspace{0.01in}
\begin{minipage}{0.47\columnwidth}
\centering
\includegraphics[width=\textwidth]{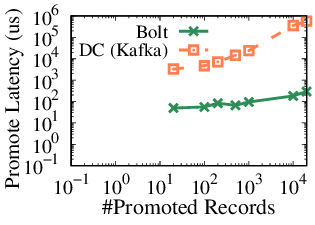}
\rcaption{fig-promote}{Promote Latency. }{\footnotesize}
\end{minipage}
\par\vspace{-0.2in}
\end{figure}

%% file: eval/adhoc-analytics/fig-analytics.tex
\begin{figure*}
\centering
\begin{minipage}{0.7\textwidth}
\centering
\subfloat[Kafka Latency over Time]{\includegraphics[width=0.35\textwidth]{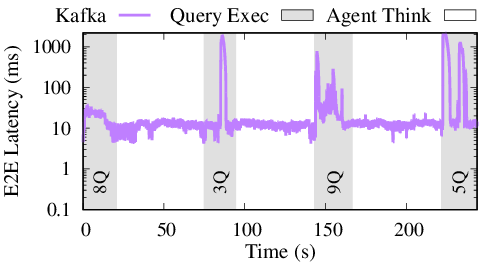}}
\subfloat[\sysname\ Latency over Time]{\includegraphics[width=0.35\textwidth]{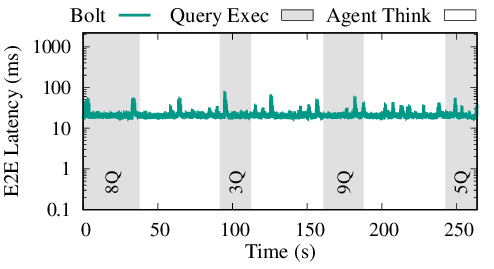}}
\subfloat[Mean and Tail Latency]{\includegraphics[width=0.3\textwidth]{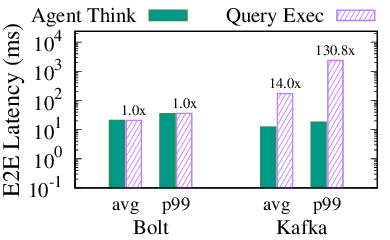}}
\vspace{-0.04in}
\rcaption{fig-analytics}{Ad-hoc Analytics Agent: Performance Isolation for a Latency-Critical Workload}{\footnotesize}
\end{minipage}
\hspace{0.01in}
\rule[-1.5cm]{0.2mm}{3.5cm}
\hspace{0.01in}
\begin{minipage}{0.24\textwidth}
\centering
\includegraphics[width=\textwidth]{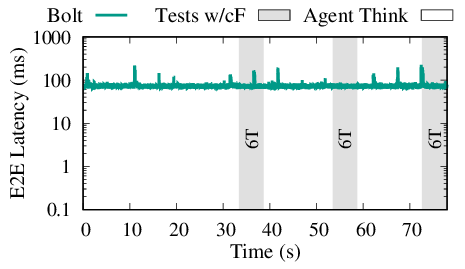}
\rcaption{fig-stream-test}{Real-time E2E Latency w/ Testing Agent. }{\footnotesize}
\end{minipage}
\par\vspace{-0.5in}
\end{figure*}

\begin{figure*}
\begin{minipage}{0.48\textwidth}
\centering
\includegraphics[width=0.49\textwidth]{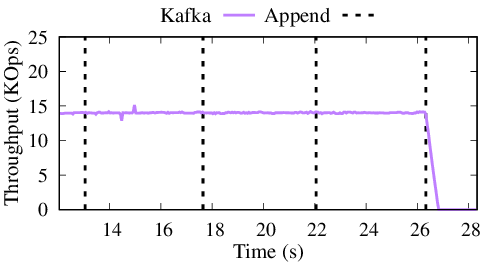}
\includegraphics[width=0.49\textwidth]{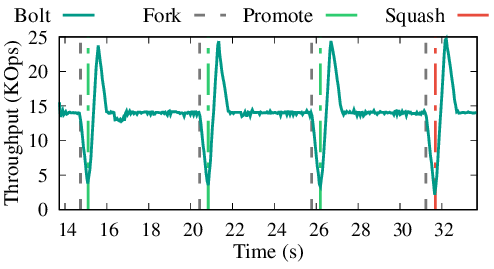}
\rcaption{fig-promote-validate}{Supply-Chain Agent: Consumer Throughput.}{\footnotesize}
\end{minipage}
\hspace{0.01in}
\rule[-1.5cm]{0.2mm}{3.3cm}
\hspace{0.01in}
\begin{minipage}{0.48\textwidth}
\centering
\subfloat[\crchange{Without Lease-based Squashing}]{\includegraphics[width=0.49\textwidth]{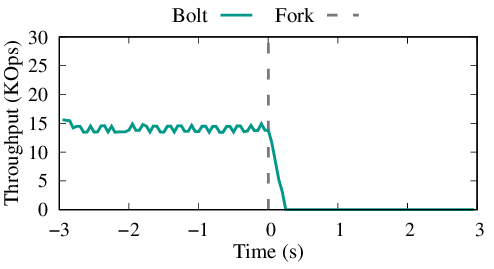}}
\subfloat[\crchange{With Lease-based Squashing}]{\includegraphics[width=0.49\textwidth]{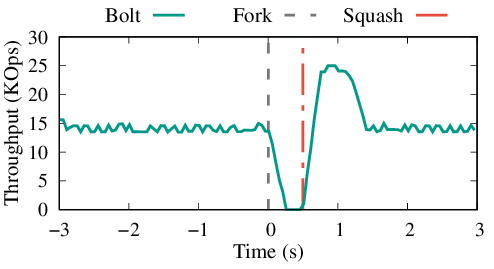}}
\rcaption{fig-rogue-agent}{\crchange{Behaviour with Rogue Agent.}}{\footnotesize}
\end{minipage}
\vspace{-0.25in}
\end{figure*}

%% file: eval/promote-validate/fig-apps.tex
% Intentionally empty; figure definitions are included elsewhere.

%% file: related.tex
\section{Related Work}
\label{sec-related}

\noindent
\textbf{Shared Logs.} Prior shared logs change the interface for virtualization~\cite{delos}, low-latency~\cite{lazylog-tocs, lazylog, speclog}, and flexible ordering~\cite{fuzzy, flexlog}. However, none provide the forking capability. While FuzzyLog~\cite{fuzzy} {\em implicitly} allows forked histories during network partitions, forking is not a first-class primitive \crchange{and is not designed} for agentic use. To our knowledge, \absname\ is the first shared log abstraction with forking as a core primitive to support agents. Like diskless designs, prior shared logs like Scalog~\cite{scalog} and Boki~\cite{boki} also separate data and metadata. However, \crchange{their storage servers are stateful (use local disks), which makes them prone to storage contention} like Kafka (\sref{sec-eval-perf-iso}). \sysname\ avoids this via its diskless design.  

\noindent
\textbf{Fork as a Core Primitive.} Beginning with OS process forks, \crchange{a fork} as a first-class operation has been well explored in many contexts~\cite{cxlfork, spork, mitosis, mufork}. Prior storage systems have also proposed forking (\crchange{sometimes referred to} as branching) primitives~\cite{aguilera2006olive, tardis}. More recently, to support AI agents, a few databases~\cite{postgres-tiger-data, neon-agent, berkeley2025overlords}, object stores~\cite{tigris-fork, tiger-fluid-storage}, lakehouses~\cite{lakehouse-agent}, and file-systems~\cite{wang2026fork} have realized the need for and adopted forking/branching as a first-class primitive. However, no shared log today provides forks. Further, unlike these systems, \absname\ provides a novel form of continuous forks.

\noindent
\textbf{Fork vs. Branch.} Branching, unlike forking, is desired when divergent paths are intended to be merged~\cite{dolt-fork-vs-branch}. A few databases such as Dolt~\cite{dolt} offer mergeable branches, but most do not~\cite{sehn2024databasebranches}. Tardis~\cite{tardis} is an earlier system that uses a branch-on-conflict mechanism with merge capability for weakly-consistent systems. \absname's promote is a restricted merge where updates from a single forked child are effectively merged into the parent. A more general notion of merge that can merge updates from multiple forks into the parent would not preserve linearizable ordering of appends. Further, our motivating use cases \crchange{did not require it.}

%% file: conc.tex
\section{Conclusion}
\label{sec-conc}

\absname\ is a new shared log that offers forking as a core primitive, enabling agents to operate over forks of a shared-log-based stream in a safe and isolated manner. As AI agents increasingly operate over data systems, it is imperative to build systems abstractions that enable safe agent interactions without interference to traditional applications. Our work takes a step in this direction for streaming data systems.

%% file: acks.tex
\vspace{0.1in}
\noindent
\textbf{Acknowledgments.} We thank our shepherd and SOSP'26 reviewers for their insightful comments. We thank Madhav Jivrajani for helping us conduct early experiments. We thank the other DASSL members and Xudong Sun for their feedback. This material was supported by funding from NSF grants CNS-2340218, CNS-2339784, IIS 2551556, IIDAI grants, as well as gifts from NetApp and TigerBeetle.

%% file: bib.tex
{	
   \bibliographystyle{plain}
   \interlinepenalty=10000
   \bibliography{all-defs,ds,urls,all-confs}
}